\documentclass[conference]{IEEEtran}
\IEEEoverridecommandlockouts
\usepackage{cite}
\usepackage{amsmath,amssymb,amsfonts}
\usepackage{algorithmic}
\usepackage{graphicx}
\usepackage{textcomp}
\usepackage{mathtools}
\usepackage{caption}
\usepackage{subcaption}
\usepackage{xcolor}
\usepackage{float}
\graphicspath{ {./images/} }

\def\BibTeX{{\rm B\kern-.05em{\sc i\kern-.025em b}\kern-.08em
    T\kern-.1667em\lower.7ex\hbox{E}\kern-.125emX}}

    \makeatletter
\newcommand{\linebreakand}{%
  \end{@IEEEauthorhalign}
  \hfill\mbox{}\par
  \mbox{}\hfill\begin{@IEEEauthorhalign}
}
\makeatother
\begin{document}
%Are footpaths encroached by shared e-scooters?
\title{ Are footpaths encroached by shared e-scooters?
Spatio-temporal Analysis of Micro-mobility Services\\}

\author{\IEEEauthorblockN{ Hiruni Kegalle}
\IEEEauthorblockA{\textit{School of Computing Technologies} \\
\textit{RMIT University}\\
hiruni.kegalle@student.rmit.edu.au}
\and
\IEEEauthorblockN{ Danula Hettiachchi}
\IEEEauthorblockA{\textit{School of Computing Technologies} \\
\textit{RMIT University}\\
danula.hettiachchi@rmit.edu.au}
\and
\IEEEauthorblockN{ Jeffrey Chan}
\IEEEauthorblockA{\textit{School of Computing Technologies} \\
\textit{RMIT University}\\
jeffrey.chan@rmit.edu.au}
\linebreakand
\IEEEauthorblockN{ Flora Salim}
\IEEEauthorblockA{\textit{Computer Science and Engineering} \\
\textit{University of New South Wales}\\
flora.salim@unsw.edu.au}
\and
\IEEEauthorblockN{ Mark Sanderson}
\IEEEauthorblockA{\textit{School of Computing Technologies} \\
\textit{RMIT University}\\
mark.sanderson@rmit.edu.au}
}

\maketitle

\begin{abstract}
Micro-mobility services (e.g., e-bikes, e-scooters) are increasingly popular among urban communities, being a flexible transport option that brings both opportunities and challenges. As a growing mode of transportation, insights gained from micro-mobility usage data are valuable in policy formulation and improving the quality of services. Existing research analyses patterns and features associated with usage distributions in different localities, and focuses on either temporal or spatial aspects. In this paper, we employ a combination of methods that analyse both spatial and temporal characteristics related to e-scooter trips in a more granular level, enabling observations at different time frames and local geographical zones that prior analysis wasn't able to do. The insights obtained from anonymised, restricted data on shared e-scooter rides show the applicability of the employed method on regulated, privacy preserving micro-mobility trip data. Our results showed population density is the topmost important feature, and it associates with e-scooter usage positively. Population owning motor vehicles is negatively associated with shared e-scooter trips, suggesting a reduction in e-scooter usage among motor vehicle owners. Furthermore, we found that the effect of humidity is more important than precipitation in predicting hourly e-scooter trip count. Buffer analysis showed, nearly 29\% trips were stopped, and 27\% trips were started on the footpath, revealing higher utilisation of footpaths for parking e-scooters in Melbourne. 

\end{abstract}


\section{Introduction}

Micro-mobility is an emerging transportation option that offers convenience and flexibility in travelling while reducing fuel consumption, traffic congestion, and environmental pollution \cite{HOSSEINZADEH2021103047, AClashofNeeds}. Many cities, including Paris, Barcelona, New York, Washington, London, and Milan have embraced micro-mobility into their urban transport system. An e-scooter sharing system is one such micro-mobility service where individuals rent e-scooters using the service provider's mobile application. Riders can scan the QR code displayed on an e-scooter to start a trip and leave it locked within the operational area at the end of the trip.

Considering the growing rate of these services, there is a need to explore the usage patterns and the spatio-temporal factors impacting the usage of micro-mobility services to inform better policies to increase usage, safety and convenience.
%On average, 5,200 e-scooter trips have been made each day within the first 17 weeks of an e-scooter trial in the City of Melbourne \cite{MelbourneScooterStat}. 
Previous work aimed at urban policy formulation showed that micro-mobility trip distribution, usage, and their association with spatio-temporal features vary based on the locality \cite{HOSSEINZADEH2021103016, BAI2020264}. For example, in Austin, US, higher e-scooter usage was measured later in the week, whereas the usage stays stable on different days of the week in Minneapolis, US. While one-million e-scooter rides were reached in Melbourne, Australia, within four months, it took nearly one year to get the same number of rides in London, UK, a city with almost twice the population and more than twice the number of shared e-scooters \cite{MelbourneScooterStat}. Therefore, a better understanding of the city-specific associations of spatio-temporal features with micro-mobility rides is needed to assist in decision making (e.g., enabling road rules, re-distributing vehicles, assigning parking points) towards service quality improvement.

Generally, the association between shared e-scooter rides and related spatio-temporal features were explored using regression models \cite{YOUNES2020308,NOLAND2021114,TOKEY2022100037,HOSSEINZADEH2021103016}. The existing methods mostly deploy a single model, which limits the ability to compare prediction accuracy and model fitness. It reduces the reliability of outputs that are interpreted based on coefficients \cite{HOSSEINZADEH2021103047, BAI2020264}. Moreover, the impact of either spatial or temporal variables were explored, providing a partial view of the spatio-temporal features associated with micro-mobility trips \cite{HOSSEINZADEH2021103016, BAI2020264, HOSSEINZADEH2021103047,NOLAND2021114}. Rigorously analysing the influence of both spatial and temporal features of the phenomenon is important to generate a more precise picture, leading to better decisions. Another limitation in the existing literature is developing an individual global model with aggregated trip data relevant to the entire study area or duration \cite{HOSSEINZADEH2021103047, HOSSEINZADEH2021103016}. It limits the capability of gaining potential insights into temporal and spatial dynamics of the relationship between micro-mobility service usage and related factors. In general, freely available micro-mobility trip data does not contain comprehensive information due to privacy concerns. Nonetheless, most findings from existing studies are based on detailed trip data, enriched with duration and distance of rides \cite{TOKEY2022100037, YOUNES2020308}. Such studies cannot be easily conducted on real-world trip data, which has limitations.

Addressing the challenges noticed in the existing work, we deployed a collection of methods to investigate the features associated with e-scooter ridership both spatially and temporally. Further, we extended the analysis to a more granular level and explored the temporal variations of spatial-feature relationships and the spatial dynamics of temporal-feature relationships. Also, a buffer analysis was conducted to gain additional insights into the relationships between Points Of Interest (POIs), and micro-mobility trip origin/destination locations in different time frames. To understand the association between spatio-temporal attributes and micro-mobility usage, we used a dataset consisting of 175,609 e-scooter trip-starts, and 180,373 trip-stops that occurred during 3 months from 1st, August 2022 in Melbourne, Australia.

Our main contributions can be summarised as follows:
\begin{itemize}
\item We provided a comprehensive analysis that helps to understand the spatial and temporal factors impacting micro-mobility services. Our analysis is built based on multiple models, therefore, generating more reliable outputs. Generally, we cover the following relations (1) E-scooter trip occurrence and spatial features. (2) E-scooter trip occurrence and temporal features. (3) POIs and e-scooter trip start/end locations
\item We explored the relationships on a multi-resolution scale, where different time frames and spatial zones were modeled. Our in-depth analysis helps to identify the dynamics and heterogeneity of spatial and temporal features associated with micro-mobility trips. 
\item Finally, we provided insights using real-world anonymized e-scooter trip data, which will assist in better decision making towards service quality improvement and policy making.
\end{itemize}

\section{Related Work}

Our work relates to the growing literature on micro-mobility services, coming from a wider range of disciplines, including transportation, law, sociology, and computer science. Prior work has studied the micro-mobility usage patterns, and their associated features, primarily in the transportation domain, where the researchers have examined the spatial and temporal determinants of micro-mobility usage separately.

\subsection{Micro-mobility services and associated spatial features}

Research has studied the spatial distribution of shared e-scooters in different cities. In Hosseinzadeh et al. \cite{HOSSEINZADEH2021103016}, the influence of factors relating to demographics, diversity, and design on e-scooter trips were evaluated. They implemented a Geographically Weighted Regression model, considering data from 159 traffic analysis zones in Louisville, KY. According to the global coefficient values of the model, 18-29 age population, male percentage, commercial land use, and mixed land use were identified as significant features related to e-scooter trips. A study conducted in Minneapolis  \cite{TOKEY2022100037} analysed the usage of e-scooters and their association with populations of different age groups, lane design, land use, and POI counts. The researchers used street segments as the geographical unit of analysis. They deployed Negative Binomial Regression (NBR) models for four different time frames on weekdays and another model for weekend trips, arguing the impact on related features vary based on the time of e-scooting. According to their findings, e-scooters are more popular among the 18-34 age group. Residential land use affects more morning and night trip generation, while commercial land use shows a higher impact on weekend and weekday-midday models. Bai and Jiao \cite{BAI2020264} compared e-scooter usage in Austin and Minneapolis, and explored whether the features affecting usage differ in the two cities. By implementing NBR models, they observed some features such as proximity to the city center, land use diversity, and transit accessibility positively correlate with e-scooter rides in both cities, while features like commercial land use and park availability were significant only in one city.

For our study, we selected spatial features from the literature \cite{HOSSEINZADEH2021103016, BAI2020264, TOKEY2022100037} and added family composition as a new feature, since it was identified as an influential factor in travel mode selection \cite{doi:10.1080/01441647.2017.1354942, CAIATI2020123}. Without being limited to a single regression model, we experimented with four machine learning models and selected the model with the least error metrics in prediction. Then the selected model was used to find most associated features with e-scooter trips. Tokey et al. \cite{TOKEY2022100037} points out the importance of implementing additional models to different time frames, since the influence on POIs and other features may be dynamic in different temporal dimensions. Seeing the temporal variation of e-scooter usage in Melbourne, we implemented models dividing trip data into 10 (5-weekday and 5-weekend) segments.  

\subsection{Micro-mobility services and associated temporal features}

When exploring the factors impacting shared e-scooter usage, researchers also looked at dynamically changing features such as weather and gasoline price. Hosseinzadeh et al. \cite{HOSSEINZADEH2021103047} investigated the influence of the day of the week, weather, and holidays for shared e-scooter and bike share trips in Louisville. They estimated the daily trip frequency using a Negative Binomial Generalized Additive model. Based on the coefficients of the model, weather features (e.g., rainfall, wind speed, mist) were associated with a decrease in the number of trips, and holiday was a factor that increases the trip count. Another study analysed the sensitivity of weather, gas prices, and holidays on the number of shared e-scooter and station-based bike share rides in Washington, D.C \cite{YOUNES2020308}. In that work, the authors implemented NBR models for both micro-mobility systems separately and compared the determinants. According to their findings, warmer weather increases the instances of trips, while humidity and precipitation negatively impact trip counts. However, they observed that e-scooter trips are less sensitive to weather changes compared to bike rides. Lesser physical effort needed and the convenience of parking for e-scooters were the possible reasons given for that observation. Holidays and gas prices showed a positive effect on both e-scooter and bike share trips.

\begin{figure}[bp]
\vspace{-1em}
\centerline{\includegraphics[width=1\columnwidth]{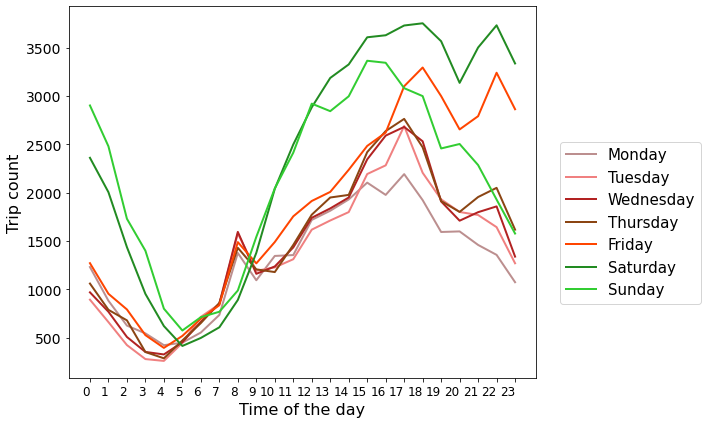}}
\caption{Hourly e-scooter trip distribution}
\label{temporalDistrbn}
\end{figure}

Since we observed sufficient variations in the number of e-scooter rides hourly as seen in Figure \ref{temporalDistrbn}, we choose to build the model with hourly granular data instead of daily. With the availability of data in Melbourne, we selected a set of weather features from the literature \cite{HOSSEINZADEH2021103047, YOUNES2020308}. In addition to the global model to estimate the number of hourly rides in the study area, we implemented three models for selected Statistical Areas. The expectation was to analyse the variation of temporal trip determinants (weather, time, day) in different geographical zones.

\subsection{Buffer Analysis}

 Tokey et al. \cite{TOKEY2022100037} and Heumann et al. \cite{su132212527} highlighted the importance of examining the presence of POIs when analysing e-scooter usage patterns. E-scooter riders participated in interviews of a study expressed that they can reach closer to their destinations with e-scooters because of the free-floating nature of the vehicle \cite{AClashofNeeds}. Li et al. \cite{LI2022101848} states trip data from dockless services such as e-scooters can be much closer to real origin and destination locations of user trips. Based on these reasons, we decided that POIs near trip start and end locations can reveal important information about usage patterns and trip intentions. Therefore, we conducted a buffer analysis to explore the impacts on POIs in different temporal dimensions.

 Buffer analysis is a common approach used in Geographic Information Systems to examine proximity \cite{GISbook}. It creates a polygon around a feature of interest that contains an area of a specified width.  Jin et al. \cite{doi:10.1080/00330124.2018.1531038} employed buffer analysis to find the spatio-temportal relationship between Uber pickup locations and public transit coverage in New York City. They analysed the number of pickups within three buffer areas around public transit stops and found that Uber both compete with and compliment public transport. In particular, their analysis on different time frames revealed competition is evident mostly in daytime, in areas with good public transit coverage, while Uber acts as a complement at midnight in areas with less public transit service. Another study used buffer analysis to explore the impact on shared e-scooters on public bus ridership \cite{ZIEDAN202120}. They implemented regression models to estimate route level bus ridership, using the variables derived from buffer analysis such as the number of e-scooter start-trips, end-trips and population within the buffer area.

Given that e-scooter trip start/end locations are closer to riders' actual origin/destination, we explored the difference between the number of trips started and ended having POIs within the buffer for multiple time frames on weekends and weekdays. Further, the number of trips started and ended within three buffer zones near paths (footpath, cycle lane, and shared path) were compared to explore the effect of e-scooter parking on different types of paths. 

\section{Datasets} 

\subsection{E-scooter trip data}

The city of Melbourne, Australia started an e-scooter trial partnered with Lime and Neuron micro-mobility service providers on 1st February 2022. Although the trial started with a fleet of 750 e-scooters for each provider, due to the growing demand, providers increased the number of e-scooters deployed in the city \cite{MelbourneScooterStat2}.

We accessed e-scooter trip details using the General Bikeshare Feed Specification API\footnote{A standardized data feed for shared mobility system availability: https://github.com/MobilityData/gbfs} provided by Lime. The \emph{free bike status} API endpoint provides a real-time snapshot of available e-scooters in Melbourne. The API response consists of a unique ID for each e-scooter, the geographical coordinates, and the battery level of the vehicle. We were able to derive trip-starts, and trip-ends by continuously collecting these API responses in regular time intervals as proposed by McKenzie \cite{10.1145/3356392.3365221}. For example, an e-scooter that appears in the available vehicle list at $t=0$, does not appear in the available list from $t=1$ to $t=3$ and reappears at $t=4$ indicates that the particular e-scooter was in use from $t=1$ to $t=3$. Hence, $t=1$ can be identified as a trip-start and $t=3$ as a trip-end.

We implemented a scheduled python script to retrieve data from the API every minute. One limitation we observed in the dataset is, the unique ID assigned to each e-scooter was changed every 15 minutes for data security purposes. Due to this limitation, we could not identify complete trips (i.e., match each trip-start with a unique trip-end) from the dataset. As a solution, we aggregated trips into 15-minute time intervals and analysed trip-starts and trip-ends separately. 
%The entire analysis was conducted based on the trip-starts and trip-ends obtained. 
Hence, the aforementioned limitation does not impact the analysis presented in this paper.

\begin{figure}[b]
\vspace{-1em}
\centerline{\includegraphics[width=1\columnwidth]{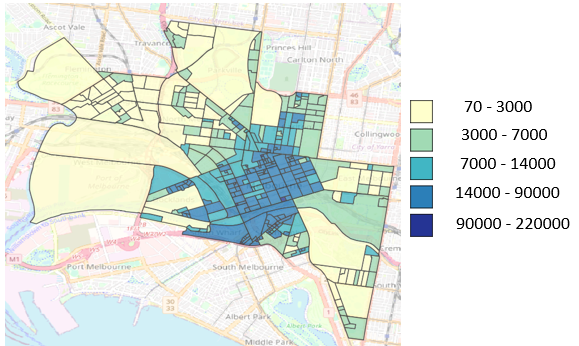}}
\caption{E-scooter trip density distribution over statistical areas}
\label{spatialDistrbn}
\end{figure}

This study used shared e-scooter trip data collected from 1st August 2022 to 30th October 2022. The city of Melbourne trial operating zone consists of 350 Statistical Areas (SA) \cite{ABS}. Figure \ref{spatialDistrbn} shows the distribution of e-scooter trip density over the study area. We presented the trip density (trips/sq. km.) instead of the trip count, since the area of different SAs are not uniform. Trip density $TD$ was calculated for each SA as follows: 
\begin{equation}
 TD (r) = {\frac{\textstyle T(r)}{\textstyle A (r)}} 
\end{equation} 
\newline Where $T(r)$ is the number of trips that occurred in region $r$ and $A(r)$ is the area of that region. We used trip-start locations, and trip-start timestamps to conduct our spatial analysis.

\subsection{Explanatory Variables of the Spatial analysis}

To find the spatial factors most associated with e-scooter trips, we modeled a set of demographic and land use variables representing each SA. These variables were selected based on the literature \cite{HOSSEINZADEH2021103016, BAI2020264, TOKEY2022100037}. Population, gender, age distribution, vehicle ownership, and family composition were collected from the platform of the Australian Bureau of Statistics \cite{ABS}. These demographic-related data for the year 2021 were available for Statistical Area 2 zones (SA2 is an aggregation of SAs). We converted these values to SA level by dividing the values by the number of SAs in each SA2 zone.

To represent the land use distribution, we collected POI data from the CLUE (Census of Land Use and Employment) project of the City of Melbourne Open Data Portal \cite{MelbourneOpenDataPortal}. It includes information about land use from 2002 and is updated annually. We selected records in census year 2020, as it was the most updated dataset. Each record contained the coordinates of POIs. Locations of the tram stops in the study area were acquired from the Open Street Map \cite{OSM}.
We calculated two design indices used in the previous studies: Shanon's entropy, which quantifies the diversity of a region, and the Mixed used Index ($MXI$), which captures the variance between residential and other POIs \cite{HOSSEINZADEH2021103016,BAI2020264}. The Shanon's entropy $E$ and the $MXI$ of a region are calculated as equations (2) and (3), respectively.

\begin{equation}
\vspace{-0.5em}
 E(r) =  \Sigma_ i \left( {\frac{\textstyle N i(r)}{\textstyle N (r)}}\times log {\frac{\textstyle N i(r)}{\textstyle N (r)}} \right)
\end{equation}
$Ni(r)$ is the POI count of type $i$ in region $r$ and $N(r)$ is the total POI count in region $r$.
\begin{equation}
\vspace{-0.5em}
 MXI(r) =  \left| {P (r) - 50} \right|
\end{equation}
$P(r)$ is the residential percentage in region $r$.
%\newline Table \ref{tabSpatial} summarises all the variables used in the spatial analysis.

\setlength{\tabcolsep}{10pt} 
\renewcommand{\arraystretch}{1.1} 
\begin{table*}[hbt!]
\caption{Descriptive statistics of dependent and independent variables of the spatial analysis}

\begin{center}
\begin{tabular}{p{1.3cm} p{2.8cm}  p{6.7cm} r r r r}
\hline
Reference & Study variable & Description & Mean & Std & Min & Max \\
\hline
\cite{HOSSEINZADEH2021103016} & Trip Density (DV)  & Total trips occurred with in a square km in a SA & 17869.9 & 21144.0 & 77.3 & 216428.5 \\
\cite{HOSSEINZADEH2021103016}, \cite{BAI2020264} & Population Density &  Number of people within a square km in a SA & 11289.0 & 8289.8 & 0.0 & 31052.4 \\
\cite{HOSSEINZADEH2021103016} & Female\% & Percentage of females in a SA & 49.6 & 4.9 & 0.0 & 52.5\\ 
\cite{HOSSEINZADEH2021103016} & Male\% & Percentage of males in a SA & 49.4 & 4.8 & 0.0 & 61.5\\ 
\cite{HOSSEINZADEH2021103016} & Age 5-14\% & Percentage of people between 5 and 14 years old in a SA & 2.8 & 2.0 & 0.0 & 7.4\\
\cite{HOSSEINZADEH2021103016} & Age 15-29\% & Percentage of people between 15 and 29 years old in a SA & 41.9 & 13.5 & 0.0 & 65.1\\
\cite{HOSSEINZADEH2021103016} & Age 30-39\% & Percentage of people between 30 and 39 years old in a SA & 25.7 & 5.7 & 0.0 & 34.0\\
\cite{HOSSEINZADEH2021103016} & Age 40-49\% & Percentage of people between 40 and 49 years old in a SA & 10.4 & 3.1 & 0.0 & 20.7\\
\cite{HOSSEINZADEH2021103016} & Age 50-64\% & Percentage of people between 50 and 64 years old in a SA & 10.0 & 4.6 & 0.0 & 20.6\\
\cite{HOSSEINZADEH2021103016} & Age 65Above\% & Percentage of people above 65 years old in a SA & 8.0 & 5.7 & 0.0 & 24.0\\
\cite{TOKEY2022100037} & Car Ownership\% & Percentage of dwellers owning motor vehicles in a SA & 55.8 & 20.1 & 0.0 & 88.0 \\
New & Without Children\% & Percentage of families with no children in a SA & 73.8 & 11.0 & 0.0 & 100.0\\
New & With Children\% & Percentage of families with  children in a SA & 25.2 & 8.9 & 0.0 & 42.5 \\
\cite{TOKEY2022100037} & Cafe\% & Percentage of cafe and restaurant in a SA & 60.6 & 40.3 & 0.0 & 100.0\\
\cite{TOKEY2022100037} & Shop\% & Percentage of retail shops in a SA & 0.5 & 2.9 & 0.0 & 50.0 \\
\cite{BAI2020264} & Office\% & Percentage of offices in a SA & 2.1 & 6.4 & 0.0 & 100.0 \\
\cite{HOSSEINZADEH2021103016}, \cite{BAI2020264} & Recreation Count &  Number of Leisure/Recreation places in a SA & 0.2 & 0.8 & 0.0 & 12.0 \\
\cite{HOSSEINZADEH2021103016} & Campus Count &  Number of universities in a SA & 0.0 & 0.1 & 0.0 & 1.0 \\
\cite{HOSSEINZADEH2021103016}, \cite{BAI2020264} & Entropy &  Entropy calculated using equation (2) & 0.6 & 0.47 & 0.0 & 2.2 \\
\cite{HOSSEINZADEH2021103016}, \cite{BAI2020264} & MXI &  MXI calculated using equation (3) & 40.5 & 13.5 & 0.0 & 50.0 \\
\cite{HOSSEINZADEH2021103016} & Tram Density & Number of trams stops within a square km in a SA & 18.2 & 40.0 &0.0 & 357.1\\
\cite{HOSSEINZADEH2021103016}, \cite{BAI2020264} & Bus Density & Number of bus stops within a square km in a SA & 20.3 & 56.7 &0.0 & 714.2 \\
\cite{HOSSEINZADEH2021103016} & Train Density & Number of metro stations within a square km in a SA & 0.2 & 1.7 & 0.0 &24.6 \\
\hline
\multicolumn{4}{l}{\tiny DV - Dependent variable}
\end{tabular}
\label{tabSpatial}
\end{center}
\end{table*}

\subsection{Explanatory Variables of the Temporal analysis}

To explore how temporal and weather factors influence trip frequency in Melbourne, we collected hourly weather data from the Australian Government Bureau of Meteorology platform \cite{BOM}. The day of the week and the time of the day were also modeled as variables. We selected these variables according to the literature.

Figure \ref{temporalDistrbn} illustrates the number of e-scooter trips started hourly on different days of the week. There is a clear variation between the number of trips that occurred on weekdays and weekends. Two noticeable peaks are visible on weekday trips, one around 8 am and the other around 5 pm. The number of trips started in the afternoon is higher than the number of trips started in the morning irrespective of the day. A similar observation of higher afternoon trips was common in a few other cities such as Minneapolis, Washington D.C. and Berlin  \cite{TOKEY2022100037, YOUNES2020308, su132212527}.

 Table \ref{tabSpatial} and Table \ref{tab-temporal} summarise all the variables used in the spatial and temporal analysis respectively.

\section{Method} 

Different to previous work where the association of either spatial or temporal features with micro-mobility services were analysed, we employed a combination of analytic methods with buffer analysis. Through this approach, we gained meaningful insights on micro-mobility usage and associated features. When exploring the features related to micro-mobility trips, we examined spatial features and temporal features separately. The main reason is, the temporal features are dynamic variables that vary with time, whereas spatial features (e.g., population, POI percentage etc.) are unchanged during the study period. Combining dynamic and static features together can reduce the predictive power of static features, that might obscure the importance of them \cite{9812627}. 

\begin{figure*}[h!]
\vspace{-1em}
\centerline{\includegraphics[width=0.95\textwidth]{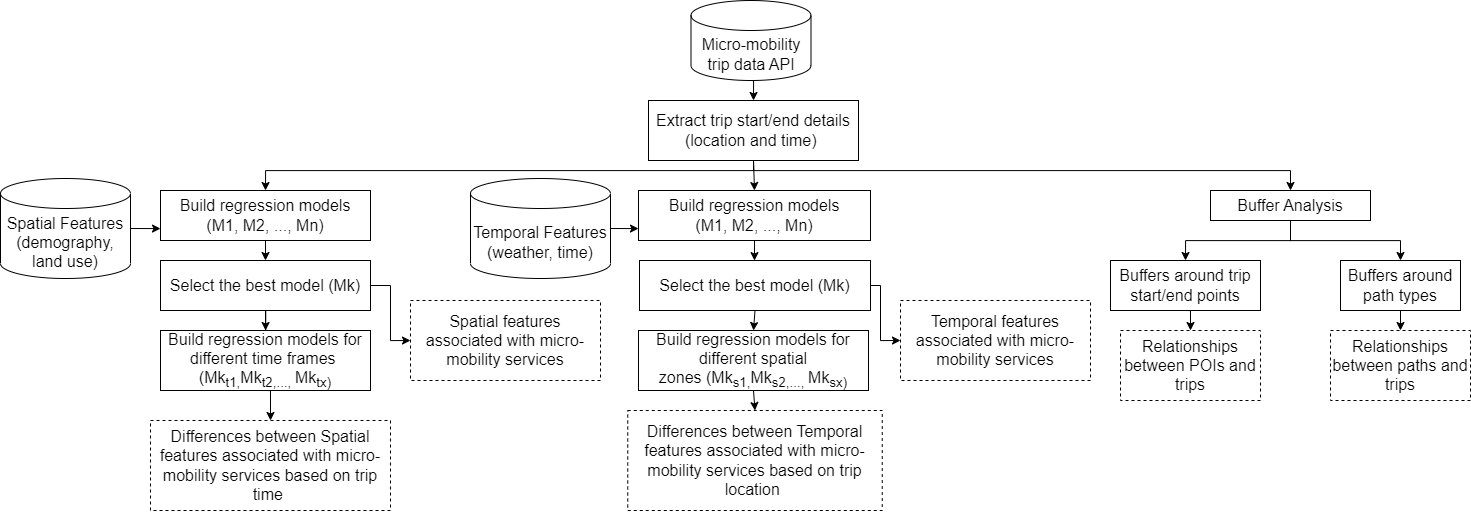}}
\caption{Methodology workflow}
\label{MethodologyWorkflow}
\end{figure*}

The proposed methodology can be easily extended to other localities and other micro-mobility trip data sets. The methodology workflow seen in Figure \ref{MethodologyWorkflow} is explained below.

\subsection{Spatial Analysis - Regression Models}

This section discusses the method employed to analyse associations between spatial features and e-scooter usage. We explored the relationship between the independent and dependent variables shown in Table \ref{tabSpatial}, using regression models. Then the analysis was extended on different time frames. Through this, we expected to investigate the temporal dynamics of the spatial features associated with e-scooter trips. 

\setlength{\tabcolsep}{10pt} % Default value: 6pt
\renewcommand{\arraystretch}{1.1} % Default value: 1
\begin{table*}[t]
\caption{Descriptive statistics of dependent and independent variables of the temporal analysis}
\begin{center}
\begin{tabular}{p{1.5cm} p{3.8cm}  p{5.7cm} r r r r}
\hline
Reference & Study variable & Description & Mean & Std & Min & Max \\
\hline
\cite{YOUNES2020308} & Hourly Trip Count (DV) & Total count of trips happened in an hour & 131.2 & 80. 2 &7.0 &453.0 \\
\cite{YOUNES2020308}, \cite{HOSSEINZADEH2021103047} & Day of week &  Day of Week & - & - & - & - \\
\cite{YOUNES2020308} & Hour of Day &  Hour of day & - & - & - & - \\
\cite{YOUNES2020308}, \cite{HOSSEINZADEH2021103047} & Humidity &  Hourly relative humidity in percentage & 75.1 & 13.5 & 37.5 & 100.0 \\
\cite{YOUNES2020308}, \cite{HOSSEINZADEH2021103047} & Precipitation &  Hourly average precipitation in mm & 0.1 & 0.2 & 0.0 & 3.8 \\
\cite{YOUNES2020308},\cite{HOSSEINZADEH2021103047} & Temperature &  Hourly average temperature in Celsius & 12.9 & 2.6 & 5.5 & 22.8 \\
\cite{YOUNES2020308}, \cite{HOSSEINZADEH2021103047} & Wind Speed &  Hourly average wind speed in mph & 5.9 & 2.7 & 0.0 & 15.6 \\
\hline
\multicolumn{4}{l}{\tiny DV - Dependent variable}
\end{tabular}
\label{tab-temporal}
\end{center}
\vspace{-6mm}%reduce white space after table
\end{table*}

\subsubsection{Regression Model for total trips}\label{AA}

We implemented four machine learning models (negative binomial regression, support vector regression, neural network, and random forest regression) to estimate the e-scooter trip density in a SA, and their performance was evaluated with error values such as MAE, MSE, RMSE, and MAPE. The best model was used to derive the features having a larger effect on the output prediction. The dataset consisting of features representing 350 SAs were randomly split into training (70\%) and testing (30\%) datasets. For all the implemented models, hyperparameters were determined using the training set and model selection was performed using the testing set.

Negative Binomial Regression (NBR) is the most commonly used modeling approach in micro-mobility literature, since it fits the overdispersed (variance exceeding mean) data better. For NBR, we used the log link function. Alpha (estimate of dispersion parameter) was calculated using an auxiliary Ordinary Least Squares regression model. As a kernel based model, we implemented Support Vector Regression (SVR) with Radial Basis kernel function. An exhaustive search procedure based on grid search was adopted for optimizing gamma and C (penalty factor for error term) parameters. The optimal settings for our model were gamma of 0.01 and C of 5. As the third model, we developed a feed forward neural network (NN) with two hidden layers. Identifying the optimal batch size and epochs of the NN was conducted using a grid search. As an ensemble regression method, Random Forest Regression (RFR) was implemented. It was optimized using grid search over a range of hyperparameters. We found that the optimal settings for our model were a 20 of estimator number, 17 of minimum sample leaf, and 10 of minimum sample split. These hyperparameters were selected based on their performance on a validation set, using the squared error as the evaluation metric. We implemented all the models with standardized features except NBR, because it is a count estimation model. 

\subsubsection{Regression Models on different time frames}

According to Figure \ref{temporalDistrbn}, there is a notable difference between the patterns of e-scooter trip occurrences on weekends and weekdays. To analyse whether there is a difference between determinants associated with trip density on weekdays and weekends, we implemented separate regression models. Then the features were compared after ordering them based on the effect they have on a model's prediction. Observing more granular temporal variations in the trip occurrence, we extended the analysis by segregating trip density data into five different time frames (0-5, 6-10, 11-15, 16-18 and 19-23) and implemented ten models (e.g., weekday\_0-5, weekend\_0-5, weekday\_6-10 etc).

\subsection{Temporal Analysis - Regression Models}

This section describes the method of analysing relationships between temporal determinants and e-scooter usage. First, hourly e-scooter trip count in the entire study area was estimated, and the most affected temporal features were derived. Then we extended the analysis on different SAs (geographical unit of analysis in this work), to explore spatial dynamics of the temporal features associated with e-scooter trips.

 \subsubsection{Regression Model for total trips}
 
We implemented NBR, SVR, NN, and RFR models to estimate the number of e-scooter trips that occurred hourly. Among the explanatory variables given in Table \ref{tab-temporal}, hourly weather data were represented as continuous variables, while day of the week (Friday as reference) and hour of day (0-5 time frame as reference) was represented as discrete variables in the model. The dataset consisting of temporal features were randomly split into training (70\%) and testing (30\%) datasets. Testing data were used to evaluate the predicting accuracy of the developed models. Parameterization of each model was completed taking the aspects discussed in section \ref{AA} into consideration. 

\subsubsection{Regression Model on different spatial zones}

Figure \ref{spatialDistrbn} shows the variation of e-scooter trip density on different SAs. To explore whether there is a difference among the temporal determinants associated with e-scooter trips in SAs, we developed separate regression models for three SAs. We selected the SAs having the highest percentage of each POI type (offices, cafes, and parks), assuming particular SAs consist of most work-related, leisure-related, and recreational trips. Then we ranked the features based on how much they contribute in each model's prediction. Through this, we aim to explore whether the gravity of temporal factors such as time of the day, day of the week and weather differ based on the trip intention. 

\subsection{Buffer Analysis}

Regression analysis helps us recognise the variables that have an impact on the number of e-scooter trips in terms of space and time. However, it does not provide insights into the association between e-scooter usage and POIs. To better understand the relationship between e-scooter parking and locations of POIs, we conducted a buffer analysis taking different time frames into account. The purpose of the buffer analysis is two-fold;

\begin{figure}[b]
\vspace{-1em}
\centerline{\includegraphics[width=1.0\columnwidth]{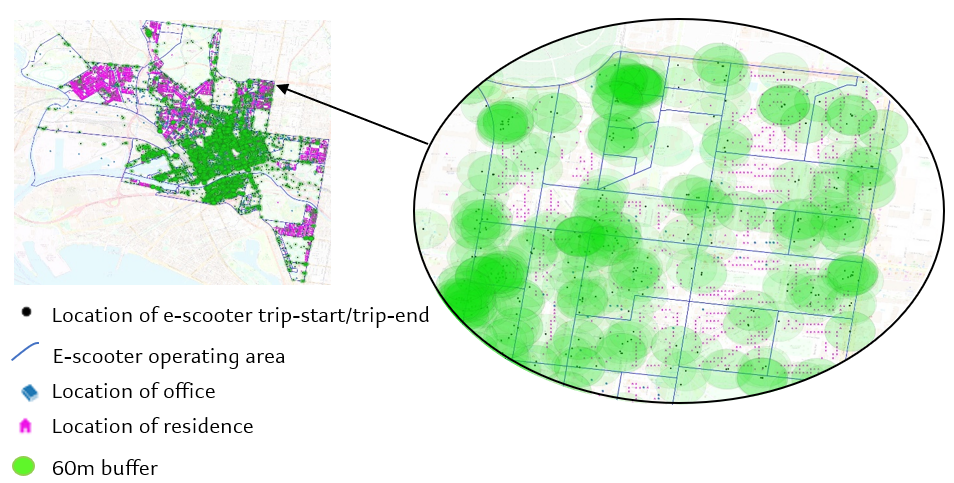}}
\caption{Buffers around e-scooter trip-start/ trip-end locations}
\label{POIbuffer}
\vspace{-1.5em}
\end{figure}

One is to explore whether there are POIs around trip start/end locations during different time frames. Figure \ref{POIbuffer} shows the placement of 60m buffer around trip-start locations and the availability of POIs. In \cite{TOKEY2022100037} a 60m buffer was used for the same purpose of capturing adjacent buildings to e-scooter parked locations. We used POI location data collected from CLUE project from the City of Melbourne. The percentage of trip start/end locations having a given type of POI within the buffer area was calculated using equation (4).

\begin{equation}
 P (x,t) = {\frac{\textstyle \Sigma_{t=i}^{j}T(x,t)}{\textstyle \Sigma_{t=i}^{j}T (t)}} 
\end{equation}

Where $T(x,t)$ is the number of trip start/end locations having at least one item of POI type $x$ inside the buffer area at time $t$ and $T(t)$ is the total number of trip start/end locations at time $t$. The value of $T(t)$ and the summary of trip start/end counts during the study period is given in Table \ref{tab-tripSummary}.

\begin{figure}[t]
\centerline{\includegraphics[width=1\columnwidth]{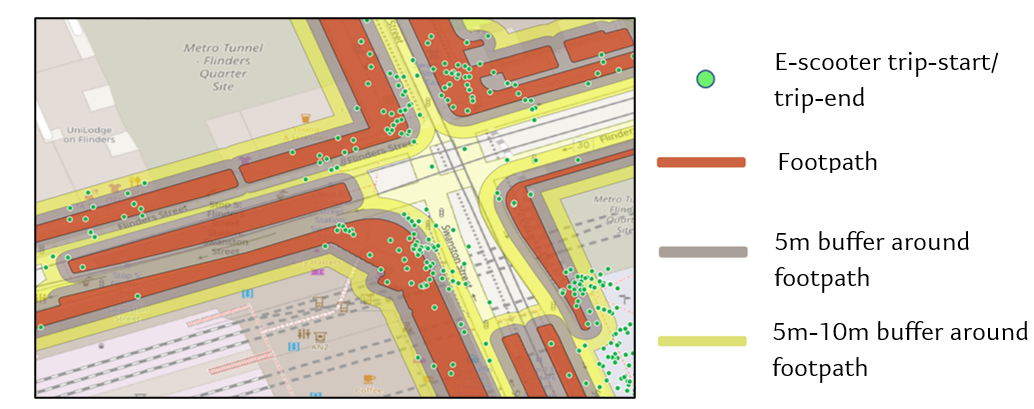}}
\caption{Buffers around footpaths}
\label{Pathbuffer}
\vspace{-1em}
\end{figure}
Our second objective of buffer analysis is to analyse the effect of e-scooter parking on different path types. For this, we collected footpath, shared path, and cycle lane shapefile datasets from the City of Melbourne open data portal. We created a 10m buffer around each path type and calculated the percentage of trips started/ended within the buffer area at different time frames using equation (4). Here, $T(x,t)$ is the number of trip start/end locations within the buffer area of path $x$ at time $t$ and $T(t)$ is the total number of trip start/end locations at time $t$. Since we observed different patterns of e-scooter parking around each path type, we extended the analysis by breaking the 10m buffer into three smaller buffers as shown in Figure \ref{Pathbuffer}; on the path, within 5m buffer, and 5m-10m buffer. We used the QGIS tool for buffer creation. 

\setlength{\tabcolsep}{7pt} % Default value: 6pt
\renewcommand{\arraystretch}{1.3} % Default value: 1
\begin{table}[b]

\caption{Summary of e-scooter trip start and end counts}
\begin{center}
\begin{tabular}{p{1.76cm}| r r r r r r}
\hline
 & \multicolumn{3}{c}{Trip Start} & \multicolumn{3}{c}{Trip Ends} \\
 
 & count & \% & per-day & count & \% & per-day \\
\hline
Weekday (WD) & 109666 & 62 & 1687.2 & 112354 & 62 & 1728.5 \\
Weekend (WE) & 65943 & 38 & 2536.3 & 68019 & 38 & 2616.2 \\
WD (0-5) & 12637 & 12 & 194.4 & 14656 & 13 & 225.48 \\
WD (6-10) & 15648 & 14 & 240.7 & 15770 & 14 & 242.62 \\
WD (11-15) & 29440 & 27 & 452.9 & 29860 & 27 & 459.38 \\
WD (16-18) & 22235 & 20 & 342.1 & 22030 & 20 & 338.92 \\
WD (19-23) & 29706 & 27 & 457.0 & 30038 & 27 & 462.12 \\
WE (0-5) & 10562 & 16 & 406.2 & 11914 & 18 & 458.23 \\
WE (6-10) & 6650 & 10 & 255.8 & 6761 & 10 & 260.04 \\
WE (11-15) & 18367 & 28 & 706.4 & 17987 & 26 & 691.81 \\
WE (16-18) & 12308 & 19 & 473.4 & 12701 & 19 & 448.50 \\
WE (19-23) & 18056 & 27 & 694.5 & 18656 & 27 & 717.54 \\
\hline
\end{tabular}
\label{tab-tripSummary}
\end{center}
\vspace{-6mm}%reduce white space after table
\end{table}

Each type of path is dedicated for a certain purpose. For example, a footpath is designated for pedestrians, a cycle lane is for cyclists and a shared path can be used by both. The aim of the extended buffer analysis around paths, is to explore how each of these path types were affected by e-scooter parking.

\section{Results}

\subsection{Spatial}\label{BB}

To select the best spatial regression model, we assessed the performance of the implemented models in terms of trip density prediction. Table \ref{tab-TotalSpatialModels} depicts the regression model results with respect to the standard error metrics; Mean Absolute Error (MAE), Mean Squared Error (MSE), Root Mean Squared Error (RMSE) and Mean Absolute Percentage Error (MAPE). We compared the standard errors for the SVR, NN, RFR models and (non-standardized) NBR, (non-standardized) RFR separately. Results show that the RFR has the least MAE, MSE and RMSE among all four machine learning models. Therefore, we selected RFR to derive the most influential features among independent variables and for the further spatial analysis of building regression models for different time frames.

\begin{figure}[bp]
\vspace{-1em}
\centerline{\includegraphics[width=1\columnwidth]{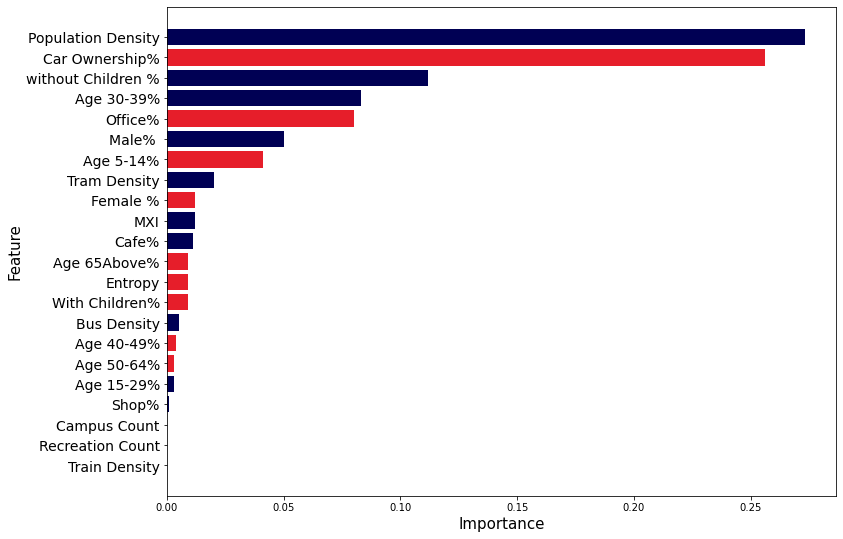}}
\caption{Feature importance of spatial regression model}
\label{RFRspatialTotal}
\end{figure}

In a regression model, we can understand the importance of features based on how they contribute to the model output prediction. We used the Scikit-learn built in \textit{feature\_importances\_} method to calculate an importance score for each independent variable. Even though this score quantifies how important each feature is for the prediction, the direction of correlation with the output is not visible. Therefore we calculated the Pearson correlation between e-scooter trip density and each feature. The feature importance scores of the spatial variables according to the RFR model is illustrated in Figure \ref{RFRspatialTotal}. Features represented in blue colour bars are positively correlated with e-scooter trip density and the red ones are negatively correlated.

\setlength{\tabcolsep}{10pt} % Default value: 6pt
\renewcommand{\arraystretch}{1.3} % Default value: 1
\begin{table}[t]
\caption{Results of regression models - spatial analysis}
\begin{center}
\begin{tabular}{p{0.5cm}| r r |r r r }
\hline
& NBR*  & RFR* & RFR & NN & SVR \\
\hline
MAE & 10539.5 & 7722.6 & 0.366 & 0.385 & 0.381 \\
MSE & 26.1e7 & 14.9e7 & 0.344 & 0.350 & 0.369 \\
RMSE & 16.1e3 & 12.2e3 & 0.578 & 0. 591 & 0.607 \\
MAPE & 0.87 & 0.89 & 2.38 & 1.91 & 2.05 \\
\hline
\multicolumn{4}{l}{\tiny* Values with non standardized variables}
\end{tabular}
\label{tab-TotalSpatialModels}
\end{center}
\vspace{-8mm}%reduce white space after table
\end{table}

As per the spatial regression model for total trips, population density in a SA is the most important feature in predicting e-scooter trip density. SAs with higher population tend to have higher e-scooter usage. Percentage of population owning motor vehicles is the second important feature according to the model. It is negatively correlated with the e-scooter trip density. The percentage of families without children is a new feature that we added in this study. Being the third important feature, family composition is an influencing factor for e-scooter trip prediction.  According to our analysis, the most important age group for predicting e-scooter trip density is 30-39 years old.

\begin{figure*}[h]
\centerline{\includegraphics[width=1\textwidth]{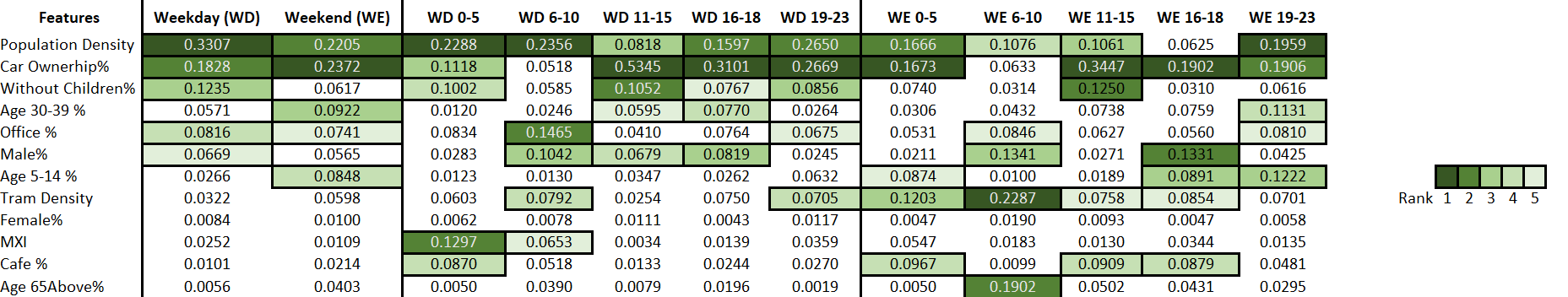}}
\caption{Feature importance - spatial analysis on different time frames}
\label{feature importance-spatial}
\vspace{-1em}
\end{figure*}

To understand the temporal dynamics of spatial features associated with e-scooter trips, we compared feature importance scores of the models implemented for different time frames. Figure \ref{feature importance-spatial} shows the feature importance scores of the top most variables of models developed for weekday and weekend different time frames. The importance of car ownership and population of young age groups increases in weekend model compared to weekday model. Population density is the most important feature in the weekday morning models and car ownership \% becomes more important in midday, evening and at night weekday models. The importance of office\% is more evident in the weekday 6-10 model. Tram stop density is an important feature in all the weekend models except 19-23. Cafe\% is important for e-scooter trip density prediction in 0-5 weekday model and three weekend models.  

\subsection{Temporal}

\begin{figure}[h]
\centerline{\includegraphics[width=1\columnwidth]{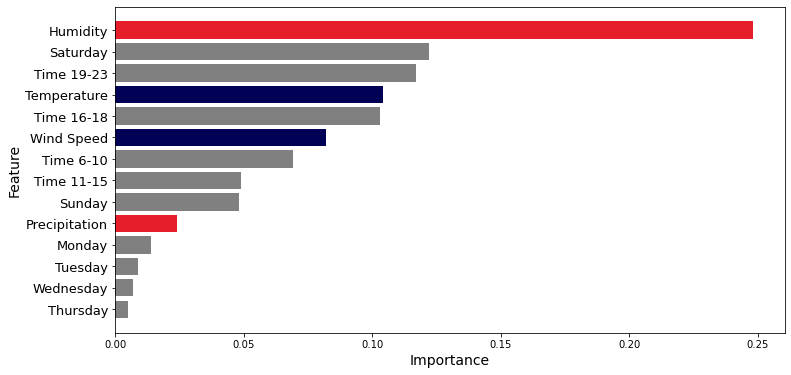}}
\caption{Feature importance of temporal regression model}
\label{temp-regression}
\vspace{-1em}
\end{figure}

The performance of temporal regression models implemented were evaluated using the metrics in section \ref{BB}. Based on the model results shown in Table \ref{tab-GlobalTemporalModels}, we selected the RFR model to explore the effectiveness of explanatory variables in trip count prediction. The same model was chosen for further temporal analysis of building regression models on different spatial zones. Figure \ref{temp-regression} depicts the feature importance scores of the temporal variables in the RFR model. Positively correlated continuous variables, negatively correlated continuous variables and discrete variables are represented in blue, red, and gray respectively. According to the temporal regression model for total trips, humidity is the most important feature that affects the output prediction. High humidity levels are associated with a lower number of e-scooter trips. While precipitation is the least important weather feature as per the model, though it also reduces the trip count. Wind speed is a feature that positive correlates with e-scooter trips. In our model, the 19-23 time frame is the most important of all time frames and among days of the week, weekend is more important than weekdays for e-scooter trip count prediction.

\setlength{\tabcolsep}{10pt} % Default value: 6pt
\renewcommand{\arraystretch}{1.3} % Default value: 1
\begin{table}[b]
\vspace{-1em}
\caption{Results of regression models - temporal analysis}
\begin{center}
\begin{tabular}{p{0.5cm}| r r| r r r }
\hline
& NBR*  & RFR* & RFR & NN & SVR \\
\hline
MAE & 40.25 & 32.6 & 0.41 &0.41 & 0.40 \\
MSE & 0.32 & 0.18 & 0.29 & 0.30 & 0.29 \\
RMSE & 56.38 & 42.50 & 0.53 & 0.55 & 0.54 \\
MAPE & 0.42 & 0.37 & 2.01 & 2.38 & 2.04 \\
\hline
\multicolumn{4}{l}{\tiny* Values with non standardized variables}
\end{tabular}
\label{tab-GlobalTemporalModels}
\end{center}
\vspace{-3.9mm}%reduce white space after table
\end{table}

\begin{figure}[b]
\vspace{-0.5em}
\centerline{\includegraphics[width=1\columnwidth]{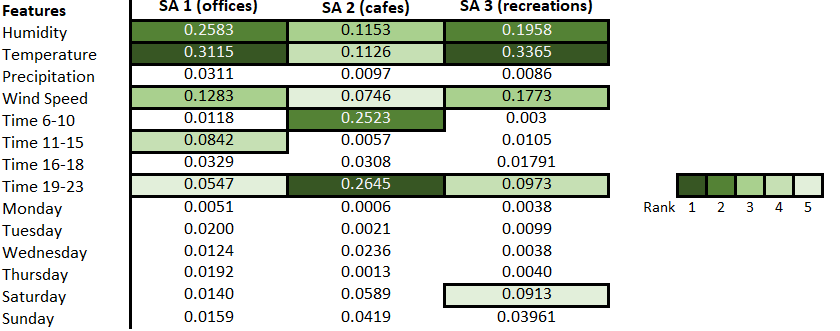}}
\caption{Feature importance - temporal analysis on different SA}
\label{featureImportance-temporal}

\end{figure}

To understand the disparity of temporal determinants in different spatial zones, we calculated feature importance scores of models implemented on three SAs. Figure \ref{featureImportance-temporal} shows the results. In the first model (SA1: with most offices), weather features show more importance than time and day features. In the second model (SA2: with most cafes), time 19-23, 6-10 are the most important  for output prediction. The importance of weather features are reduced in the second model. In the third model (SA3: with most recreation), weather features hold more importance than day and time. Under the assumption that SA1, SA2, and SA3 consist of most work-related, leisure-related and recreational trips, this results demonstrate the impact of weather on shared micro-mobility  usage vary based on the trip intention. In model 2 and 3, the important scores for Saturday and Sunday are higher than weekdays, while in model 1, some weekdays have higher important scores than weekend.
It indicates that weekends are more important in the areas with leisure-related trips. Overall, the importance of temporal variables for shared e-scooter trip count prediction is diverse across geographical areas.

\subsection{Buffer Analysis}

\begin{figure}[b]
\centerline{\includegraphics[width=1\columnwidth]{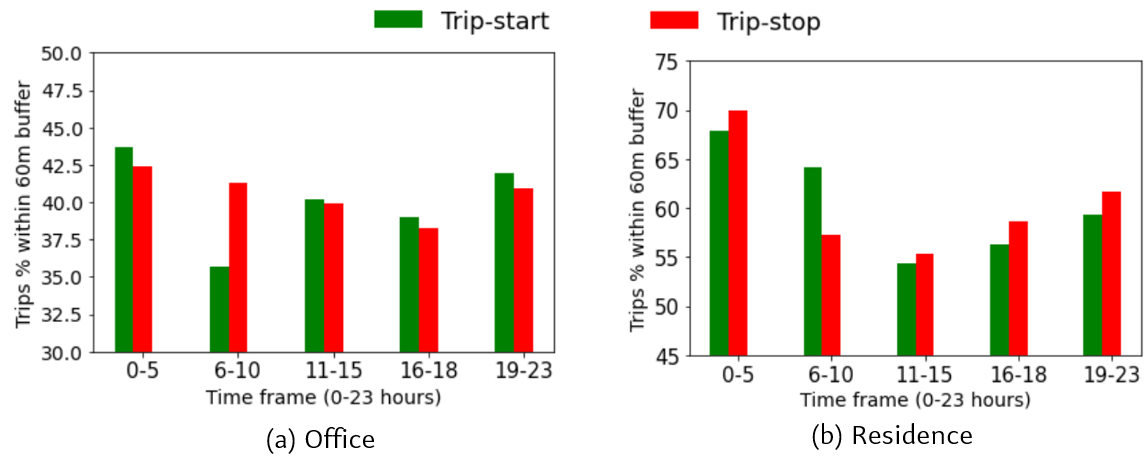}}
\caption{Percentage of trips having POIs within buffer}
\label{fig:POI60m buffer}
\vspace{-0.5em}
\end{figure}

In this section, results of the buffer analysis is presented. We selected residences and offices as the POIs of concern, since those were the most important according to the total trip regression model. Figure \ref{fig:POI60m buffer} shows percentage of trips with offices and residences within the buffer area at different time frames on weekdays.
Comparing the percentage of trip-starts having offices within the buffer area against relevant trip-stops shows, 6-10 is the only time frame having more stops (mean = 41.3\%) than starts (mean = 35.7\%). The difference between percentage of trip starts and ends having offices in the buffer area were statistically significant, confirmed by paired t-test at 0.05 significance level; t=-11.78, p$<$0.05. In contrast, in the 6-10 time frame, more trip-starts (mean = 64.2\%) happened near residences than trip-ends (mean = 57.3\%). A paired t-test verified the difference between percentage of trip starts and ends with residences in the buffer area to be statistically significant at 0.05 significance level; t=10.45, p$<$0.05.  One possible reason for these observations can be more riders start e-scooter trips around residences and reach offices in 6-10.  

\begin{figure}[t]
\centerline{\includegraphics[width=1\columnwidth]{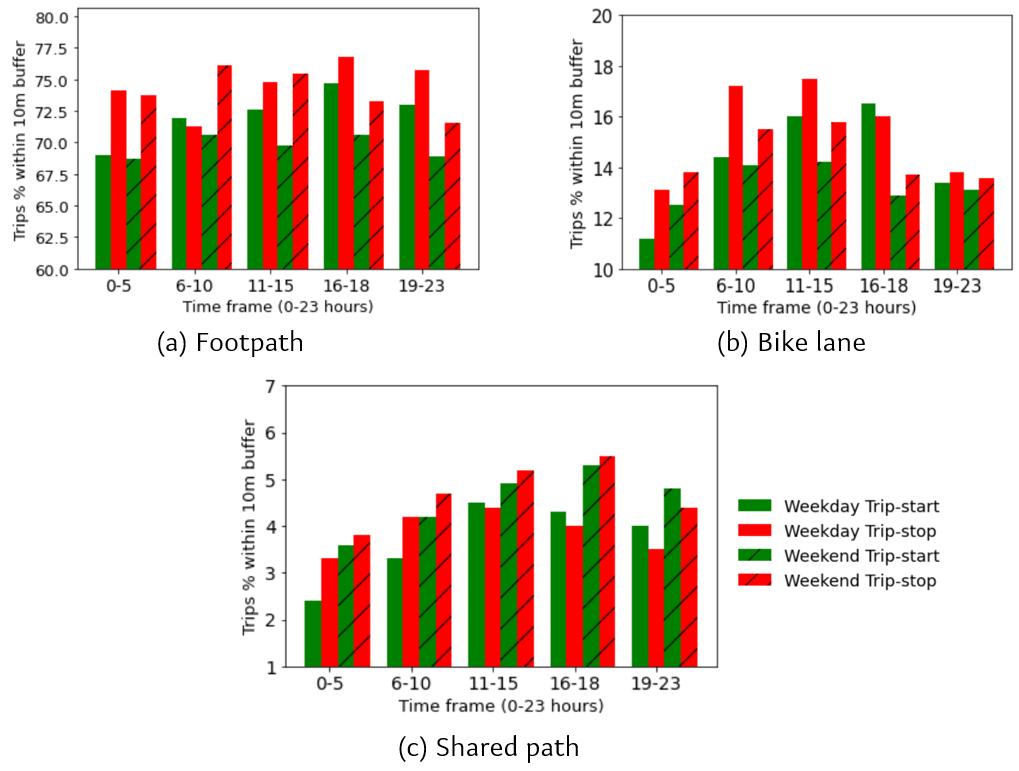}}
\caption{Percentage of trips start/end around paths}
\label{fig:path10m buffer}
\vspace{-1em}
\end{figure}

Next, we discuss the results of the buffer analysis around paths. We first looked at the percentage of trips started/ended within a 10m buffer area from each type of path. To understand e-scooter parking patterns near different path types, we compared the percentage of trips originated and ended near a selected path type on different time frames. One important thing to note here is, we did not compare the percentage of trips across path types, because the land coverage of footpaths is higher than the land coverage of cycle lanes and shared paths. The percentages of e-scooters trips originated and ended near footpaths, bike lanes and shared paths at different time frames in weekdays and weekends are shown in Figure \ref{fig:path10m buffer}. In all time frames, the percentage of e-scooters stopped near footpaths (mean = 74.4\%) are higher than the trips started (mean = 71.4\%) around a footpath. A paired t-test confirmed that difference between percentage of stopped and started e-scooter trips around footpaths were statistically significant at 0.05 significance level; t=-5.26, p$<$0.05. When comparing e-scooters parked near bike lanes, higher percentages of trip starts and stops can be seen at daytime (6-18\_mean = 15.3\%, 19-5\_mean = 13\%). Most of the shared paths in Melbourne are situated around recreation locations such as parks. It can be the reason for having more trip-starts and trip-ends near shared paths on weekends compared to weekdays in every time frame (weekday\_mean = 3.8\%, weekend\_mean = 4.6\%).

We extended the buffer analysis around paths by breaking the 10m buffer into smaller areas. Through this, we wanted to explore whether e-scooters are parked 'on the path' or around the path. We calculated the percentage of trips started and ended within three buffer areas in five time frames (similar time frames used in the spatial regression).
Table \ref{tab-pathResult} shows the mean percentage values relevant to trip-starts and trip-stops of weekdays and weekends. According to the results, parking patterns around footpaths is different to cycle lanes and shared paths. More trips have started and ended on the footpath (28.2\%) compared to the area around footpaths (5m = 25.3\%, 5m-10m = 19.1\%). The opposite is observed for cycle lanes and shared paths. It indicates the higher utilization of footpaths for parking e-scooters.

%\begin{figure}[t]
%\centerline{\includegraphics[width=0.85\columnwidth]{pathTypes.PNG}}
%\caption{Land coverage of path types in the e-scooter trial area}
%\label{pathTypes}
%\end{figure}

%\begin{figure*}[t]
%\centerline{\includegraphics[width=1.0\textwidth]{path result table.PNG}}
%\caption{Results: buffer analysis (on the paths, 5m, 5m-10m)}
%\label{pathResultTable}
%\end{figure*}

\setlength{\tabcolsep}{10pt} % Default value: 6pt
\renewcommand{\arraystretch}{1.2} % Default value: 1
\begin{table*}[t]
\caption{Percentage of trips started/ended near paths}
\begin{center}
\begin{tabular} {p{1.8cm}| p{1cm}  p{1cm} p{1cm}| p{1cm}  p{1cm} p{1cm} | p{1cm}  p{1cm} p{1cm}}
\hline
 &
\multicolumn{3}{c|}{Footpath} & 
\multicolumn{3}{c|}{Cycle lane} &
\multicolumn{3}{c}{Shared path}\\
 
  &  on path & 5m & 5m-10m & on path & 5m & 5m-10m & on path & 5m & 5m-10m\\
\hline
weekday-start  & 27.2\% & 25.4\% & 19.6\% & 1.4\% & 6.2\% & 6.6\% & 0.4\% & 1.8\% & 1.6\% \\
weekend-start  & 26.8\% & 24.8\% & 18.6\% & 1.0\% & 5.8\% & 6.4\% & 1.0\% & 2.2\% & 2.0\% \\
weekday-stop & 29.6\% & 25.4\% & 19.4\% & 1.6\% & 6.6\% & 7.2\% & 0.6\% & 2.0\% & 1.6\% \\
weekend-stop & 29.2\% & 25.8\% & 18.8\% & 1.4\% & 6.4\% & 7.0\% & 1.0\% & 2.2\% & 2.0\% \\
\hline
\end{tabular}
\label{tab-pathResult}
\end{center}
\vspace{-6mm}%reduce white space after table
\end{table*}

\section{Discussion}
The findings of this study have several novelties and are aligned with literature. Shared e-scooters are mostly used in highly populated areas, while the percentage of motor vehicle owners negatively associate with the e-scooter usage. This indicates the resistance of private motor vehicle owners to transfer into active transport modes. This study found that e-scooter trips are more likely to be generated in areas with a higher population of 30-39 years older adults. It slightly varies with the understanding of other studies on age composition and e-scooter usage, where Hosseinzadeh et al. \cite{HOSSEINZADEH2021103016} and  Tokey et al. \cite{TOKEY2022100037} found the percentage of 18-29 and 18-34 years old population is significant for e-scooter usage in Louisville, US and Minneapolis, US respectively. Our finding, the importance of cafe\% for e-scooter trip density prediction in three weekend models, is consistent with \cite{TOKEY2022100037}, which showed the significance of food related POIs in the weekend. The results of our study suggest that tram stop density is associated more with weekend e-scooter trips. It indicated the possible use of shared e-scooters to get connected to public transit. Further research in this direction is needed to validate the findings.

This study also found that high humidity levels and precipitation are associated with a lower number of e-scooter trips, and a similar observation was reported in other cities \cite{HOSSEINZADEH2021103047,NOLAND2021114, YOUNES2020308}. Although wind speed is a feature that reduced trip occurrence in some studies \cite{HOSSEINZADEH2021103047, NOLAND2021114}, it shows a positive correlation with e-scooter trips in our case. Having a lesser maximum wind speed reported in our study duration compared to the other studies can be a possible reason for this observation. The comparison of trip occurrence on different spatial zones show that trips occurring in areas with more cafes are less sensitive to weather features.

These findings will help city planners and e-scooter providers to determine the areas of future service expansion and infrastructure development. In terms of infrastructure, we found that e-scooters are likely to be parked on the footpath, which might be distracting pedestrians, specially on narrow sidewalks. Imposing appropriate limitations on e-scooter parking and improving required infrastructure to encourage safe use of e-scooters are crucial steps towards sustainably integrating micro-mobility into the urban community.

\section{Conclusion}

In this research, we provided an analysis that helps to understand the spatial and temporal factors impacting micro-mobility services, on a multi-resolution scale. We generated insights that are novel and in line with existing literature, using real-world anonymised e-scooter trip data. Our findings showed the importance of spatial features and temporal features towards predicting e-scooter trip occurrence changes over time and space. While these findings assist in service quality improvement, this method can  be  easily  extended  to  other  cities and other micro-mobility trip datasets as long as the data contains trip start/end geo-locations and time stamps.

One limitation of our work is the available data, where we conducted the analysis based on three-months of trip data related to one e-scooter provider in a single city. Even though, the analysis does not necessarily generate a full picture due to this limitation, the method we employed can be extended to incorporate data collected over a long period from multiple cities and different e-scooter providers.

\section*{Acknowledgment}
This research was conducted by the ARC Centre of Excellence for Automated Decision-Making and Society (project number CE200100005), and funded by the Australian Government through the Australian Research Council. We gratefully acknowledge the City of Melbourne for supporting us with the initial advice and Lime micro-mobility for providing access to e-scooter data in Melbourne.

%  the support of ARC Centre of Excellence for Automated Decision-Making and Society (ADM+S), and Lime micro-mobility company. 

%\clearpage,  

\bibliography{SpatioTemporalAnalysisMDM23.bib}{}
\bibliographystyle{IEEEtran}

\end{document}